\documentclass[prl,twocolumn,superscriptaddress,showpacs]{revtex4}

\makeatletter

\begin{document}
\title{Spectrum of magnetohydrodynamic turbulence}
\author{Stanislav Boldyrev}
\affiliation{Department of Astronomy and Astrophysics, University of Chicago, 
5640 S. Ellis Ave, Chicago, IL 60637;~{\sf boldyrev@uchicago.edu}}
\date{\today}
\input psfig.sty

\begin{abstract}
We propose a phenomenological theory of strong incompressible
magnetohydrodynamic turbulence in the presence of a strong large-scale 
external magnetic field. We argue that
in the inertial range of scales, magnetic-field and velocity-field
fluctuations tend to align the directions of their polarizations.  
However, 
the perfect alignment cannot be reached, it is precluded
by the presence of a constant energy flux over scales. As a consequence, 
the directions of fluid and magnetic-field fluctuations at each 
scale~$\lambda$ become effectively aligned 
within the angle~$\phi_{\lambda}\propto \lambda^{1/4}$, which leads 
to scale-dependent depletion of nonlinear interaction and to the 
field-perpendicular energy spectrum~$E(k_{\perp})\propto k_{\perp}^{-3/2}$.
Our results may be universal, i.e., 
independent of the external magnetic field, 
since small-scale fluctuations locally experience a strong field produced
by large-scale eddies.

\end{abstract}
\maketitle

{\bf 1}. {\em Introduction.}--- Magnetohydrodynamic (MHD) turbulence is pervasive in astrophysical systems, where ranges of scales 
available for plasma fluctuations span many orders of magnitude, and the fluctuations 
commonly possess power-law spectral distributions, e.g.,~\citep{biskamp}. 
For example, the spectrum 
and structure of MHD fluctuations are relevant for the physics of solar wind, interstellar 
scintillation, cosmic-ray propagation in galaxies, and heat conduction in cooling flows 
in galaxy clusters. 

The spectrum of MHD turbulence was first addressed 
by \citet{iroshnikov} and \citet{kraichnan}, who proposed the physical framework 
for describing the turbulent energy cascade mediated by a 
guiding magnetic field. However, recent numerical and analytic works have 
challenged these standard results and revived substantial interest to the 
fundamentals of strong MHD 
turbulence~\citep{goldreich,cho,biskamp3,milano,maron,biskamp-muller,muller,haugen,ng,galtier3}.  
To formulate the problem and to set the notation,  
we first describe the Irosnikov-Kraichnan~\citep{iroshnikov,kraichnan} and 
Goldreich-Sridhar~\citep{goldreich} theories, and  
point out some discrepancies of these theories with recent high-resolution 
numerical findings. 
Then, we propose a new model for MHD turbulence, which is free of such 
discrepancies, and which explains the results of 
numerical simulations~\citep{maron,biskamp-muller,muller,haugen}. 

Consider a conducting fluid stirred by a random force with the correlation 
length~$\Lambda_0$. The system size is larger 
than~$\Lambda_0$, and viscosity and resistivity of the fluid are very small. The goal is to find  
the stationary energy spectrum of the resulting turbulent fluctuations in the inertial 
interval of scales,~$\lambda\ll \Lambda_0$. Let us split the magnetic field into two 
parts, ${\bf B}(x,t)={\bf B}_0+{\bf b}(x,t)$, where ${\bf B}_0=\langle {\bf B} \rangle$  
is the system-size averaged magnetic field, and ${\bf b}(x,t)$ is the fluctuating part. 
The MHD equations describing the evolution of the magnetic field and of the 
fluid-velocity field ${\bf v}(x,t)$ can be represented in the 
so-called Els\"asser variables, ${\bf z}={\bf v}-{\bf b}$, and ${\bf w}={\bf v}+{\bf b}$:
\begin{eqnarray}
\partial_t {\bf z}+ ({\bf V}_A\cdot \nabla){\bf z}+({\bf w}\cdot \nabla){\bf z}=-\nabla P, 
\label{mhd1} \\
\partial_t {\bf w}-({\bf V}_A\cdot \nabla){\bf w}+({\bf z}\cdot \nabla){\bf w}=-\nabla P, 
\label{mhd2}
\end{eqnarray}
where ${\bf V}_{A}={\bf B}_{0}/\sqrt{4\pi \rho}$ 
is the Alfv\'en velocity,  $\rho$ is the fluid density, $P$ is the pressure 
that is determined from the incompressibility condition, $\nabla \cdot {\bf z}=0$ 
or $\nabla \cdot {\bf w}=0$, and we omit the terms representing large-scale forcing  
and small viscosity and resistivity.  

To present the standard arguments of Iroshnikov and Kraichnan, let us note that 
owing to the  symmetric form of system~(\ref{mhd1}, \ref{mhd2}) two classes of exact solutions 
exist. For ${\bf w}\equiv 0$, any function ${\bf z}={\bf g}({\bf r}-{\bf V}_A t)$ is the solution 
of the system; analogously, 
for ${\bf z}\equiv 0$, the solution is given by an arbitrary function ${\bf w}={\bf h}({\bf r}+{\bf V}_A t)$.  
From the form of the nonlinear terms in system~(\ref{mhd1}, \ref{mhd2}), one observes that Alfv\'en-wave packets, 
or ``eddies,'' propagating in the same direction along~${\bf B}_0$ do not interact. One has therefore to investigate 
interactions or eddies propagating in  opposite directions. 

Consider a wave packet of 
size $\lambda$ propagating along the large-scale field~${\bf B}_0$. We  
denote the corresponding perturbations (i.e., typical variations across the eddy) 
of the velocity and magnetic fields by 
$\delta v_{\lambda}$ and $\delta b_{\lambda}$; in the Alfv\'en 
wave, $\delta v_{\lambda}\sim \delta b_{\lambda}$. Its interaction with the 
counter propagating packet of the same size occurs during time~$\lambda/V_A$.  
As follows from (\ref{mhd1},\ref{mhd2}), during 
one interaction the eddies are deformed only slightly, 
$\Delta \delta v_{\lambda}\sim (\delta v_{\lambda}^2/\lambda)(\lambda/V_A)$. 
Since different eddies 
are not correlated, the perturbations add up randomly, so the eddy is deformed 
considerably only after a large number of interactions, $N\sim (\delta v/\Delta \delta v_{\lambda})^2$.
The time of energy transfer to a smaller eddy  
can thus be estimated as $\tau_{IK}(\lambda)\sim N\lambda/V_A\sim  \lambda/\delta v_{\lambda}(V_A/\delta 
v_{\lambda}).$ This time is larger than the Kolmogorov dynamic time, 
$\tau(\lambda)\sim \lambda/\delta v_{\lambda}$, by the Alfv\'en factor~$V_A/\delta v_{\lambda}$. 
Assuming that the energy flux 
over scales is constant, $\delta v^2_{\lambda}/\tau_{IK}={\rm const}$, we obtain the 
Iroshnikov-Kraichnan energy spectrum, 
\begin{eqnarray}  
E_{IK}(k)= \langle |\delta {\bf v}(k)|^2\rangle k^2\propto k^{-3/2}.
\label{eik}
\end{eqnarray} 

The essential assumption of the Iroshnikov-Kraichnan picture is that the eddy size is the same in 
the field-parallel and 
field-perpendicular directions. However, numerical and observational data accumulated for 
the last 30 years indicate that in MHD turbulence the energy transfer occurs predominantly 
in the field-perpendicular direction, e.g.,~\citep{milano,biskamp}. 
This raises the question whether anisotropy 
is crucial for the energy cascade, and whether it changes 
the spectrum of turbulence. 

\begin{figure} [tbp]
\vskip-10mm
\centerline{\psfig{file=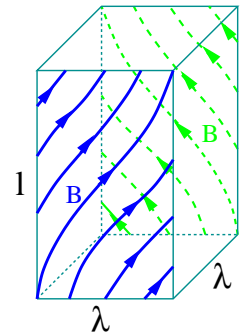,width=3.3in,angle=-90}}
\vskip-10mm
\caption{Sketch of a turbulent eddy in Goldreich-Sridhar
picture. The large-scale magnetic field is in the vertical direction. The field perpendicular dimensions  
of the eddy are the same, while its field-parallel scale is~$l\propto \lambda^{2/3}$.  
As the turbulent cascade proceeds toward the smallest,
dissipative scales,~$\lambda \to 0$, the Goldriech-Sridhar eddy assumes the
shape of a filament.}
\label{gseddy}
\end{figure}

An elegant  
treatment of anisotropic MHD turbulence was proposed by~\citet{goldreich}. They 
suggested that as the energy cascade proceeds to smaller scales, turbulent eddies 
progressively become elongated along the large-scale field. 
Their field-parallel and 
field-perpendicular scales are found from the so-called 
critical-balance condition. This condition follows from two different estimates that are 
equivalent in the Goldreich-Sridhar picture. First, the field-parallel 
scale of an eddy is found from formal balance of the linear and nonlinear terms 
in the MHD equations (\ref{mhd1},\ref{mhd2}),  
$V_A/l\sim \delta v_{\lambda}/\lambda$. Second, the field-parallel scale of an eddy 
can be obtained from the requirement that the magnetic field-line displacement in 
the eddy, $\xi \sim \delta b_{\lambda} l/V_A$, be 
comparable with the field-perpendicular eddy size,~$\lambda$. 
The shape of the turbulent eddy in the Goldreich-Sridhar theory 
is schematically presented in Fig.~\ref{gseddy}. 
As a result, two counter propagating eddies are deformed strongly during 
only one interaction, and the energy 
transfer time is given by the Alfv\'en crossing time, 
$\tau_{GS}\sim l/V_A\sim \lambda/\delta v_{\lambda}$. The Goldreich-Sridhar 
theory thus predicts that due to local anisotropy, the energy-transfer time is reduced 
to the Kolmogorov estimate.  
The field-perpendicular energy spectrum is obtained from the condition of constant energy flux,  
$\delta v_{\lambda}^2/\tau_{GS}={\rm const}$, which gives  
\begin{eqnarray}
E_{GS}(k_{\perp}) =\langle |\delta {\bf v}(k_{\perp})|^2 \rangle k_{\perp}\propto k_{\perp}^{-5/3}, 
\label{egs}
\end{eqnarray}
where $\delta {\bf v}(k_{\perp})=\int \delta {\bf v}({\bf x_{\perp}})
\exp(-i{\bf k}_{\perp}\cdot {\bf x}_{\perp})d^2{x}_{\perp}$.

Recent high-resolution numerical simulations of MHD turbulence in a strong external magnetic 
field indeed confirmed the elongation of turbulent fluctuations 
along the large-scale magnetic-field~\citep{milano,cho,maron,biskamp-muller}.  
However, the field-perpendicular energy spectrum was consistently found to be close 
to~$E(k_{\perp}) \propto k_{\perp}^{-3/2}$~\citep{maron,biskamp-muller,muller,haugen}.  
Obviously, such a spectrum combined with the anisotropy of fluctuations 
contradicts both the Iroshnikov-Kraichnan and the Goldreich-Sridhar 
phenomenologies. This controversy motivated our interest in the problem. 

In this paper we argue that filament-like 
eddies are, in fact, non-realizable.  We propose that 
the small-scale turbulent eddies spontaneously develop angular alignment 
of their  magnetic-field and velocity-field polarizations, which leads to 
their local anisotropy in the field-perpendicular plane. This effect 
is similar to the dynamic alignment known in the case of decaying MHD 
turbulence, where magnetic and velocity fluctuations approach 
the configuration ${\bf v}(x)\equiv {\bf b}(x)$ or ${\bf v}(x)\equiv -{\bf b}(x)$,  
depending of the initial 
conditions~\citep{dobrowolny,grappin,pouquet}. 
In the aligned state, the nonlinear interaction is zero, see Eqs.~(\ref{mhd1},\ref{mhd2}). 

We propose that in the case of driven turbulence the tendency to  
dynamic alignment is preserved, however, the precise alignment cannot be 
reached. The reason is an energy cascade toward small scales, which should be 
maintained by nonlinear interaction. We thus argue that 
at each scale~$\lambda$, 
the alignment of fluctuations should reach the maximal level consistent with 
a constant energy flux through this scale.   
We demonstrate that this is achieved when the velocity and magnetic-field 
fluctuations~$\delta {\bf v}_{\lambda}$ and~$\pm \delta {\bf b}_{\lambda}$ align 
their directions within the angle~$\phi_{\lambda}\propto \lambda^{1/4}$.     
The dynamic alignment in driven turbulence thus becomes scale-dependent. 
Quite remarkably, this leads to the field-perpendicular 
energy spectrum~$E(k_{\perp}) \propto k_{\perp}^{-3/2}$, which explains the numerical observations 
and resolves the above mentioned controversy. 

As another important result, in our theory small-scale eddies can be 
viewed as sheets or  ``ribbons'', stretched along the magnetic-field lines. 
This explains the well known numerical fact that the dissipative structures 
in MHD turbulence are micro current sheets rather than filaments, 
e.g.,~\citep{biskamp,biskamp3,maron}.  
In the next section we introduce our model of anisotropic MHD turbulence. Preliminary results 
on the dynamic alignment in driven MHD turbulence can be found in our earlier work~\citep{boldyrev}. 
~\\

{\bf 2}. {\em Structure and spectrum of MHD turbulence.}--- As one can check, the MHD equations~(\ref{mhd1},\ref{mhd2}) conserve 
the integrals $\int z^2d^3 x$ and $\int w^2 d^3 x$, if the fluctuations~${\bf w}(x)$ 
and~${\bf z}(x)$ have periodic 
boundary conditions or vanish at infinity. These integrals can be expressed through the integral 
of energy  
\begin{eqnarray}
E=\frac{1}{2}\int (b^2+v^2)d^3 x,
\end{eqnarray}
and the integral of cross-helicity,
\begin{eqnarray}
H^C=\int ({\bf v}\cdot {\bf b})d^3 x.
\end{eqnarray}
In the unforced case, both integrals decay due to small viscosity and resistivity of the fluid. 
However, 
dissipation of cross-helicity is not sign-definite, and, therefore, the integral of 
cross-helicity decays slower than the integral of energy, e.g.,~\citep{biskamp}.  
As a result of such ``selective decay,'' turbulence approaches the perfectly aligned 
configuration~${\bf b}(x)\equiv{\bf v}(x)$ or~${\bf b}(x)\equiv -{\bf v}(x)$ depending on the 
initial conditions. This behavior is known as the dynamic alignment 
or the Alfv\'enization effect~\citep{dobrowolny,grappin,pouquet}. In the 
aligned state, either~${\bf z}(x)$ or~${\bf w}(x)$ is identically zero and nonlinear 
interaction vanishes. 

We propose that a similar effect is present in driven 
MHD turbulence, since the external force locally produces large-scale 
fluctuations of cross-helicity, which are then inherited by smaller-scale 
eddies. Both~$E$ and $H^C$ cascade toward small scales, however, 
the cascade rate of 
cross-helicity may generally be smaller than that of energy, which forces 
fluid and magnetic fluctuations to align their polarizations at each given 
scale. However, the precise alignment cannot be reached, it would be inconsistent 
with the constant energy flux over scales. Instead, the alignment of fluctuations 
should saturate at the maximal level that can be achieved in the presence of such a flux. 

Let us first describe the shape of the eddy, which would be dictated solely 
by a constant energy flux, without any constraints imposed by the 
cross-helicity conservation (this derivation was first 
proposed in~\citep{boldyrev}). Assume that directions of 
shear-Alfv\'en velocity- and magnetic-field   
fluctuations $\delta{\bf v}_{\lambda}$ and $\pm \delta {\bf b}_{\lambda}$ 
are aligned within some (small) angle $\theta_{\lambda}$ in the 
field-perpendicular plane. 
As one can directly check, this leads to depletion 
of the nonlinear interaction 
in Eqs.~(\ref{mhd1},\ref{mhd2}): $({\bf w}\cdot \nabla){\bf z}\sim ({\bf z}\cdot \nabla){\bf w} 
\sim \theta_{\lambda}\delta v_{\lambda}^2/\lambda $. Similarly to the Goldreich-Sridhar critical balance, 
the eddy elongation in the field-parallel 
direction is found from balancing the linear and nonlinear terms in Eqs.~(\ref{mhd1},\ref{mhd2}), 
$l\sim V_A \lambda/(\delta v_{\lambda} \theta_{\lambda})$. The energy transfer time is then calculated 
as the Alfv\'en crossing time, $\tau_N\sim l/V_A\sim \lambda/(\delta v_{\lambda}\theta_{\lambda})$. It is 
important that such turbulence is strong and essentially three-dimensional. 

\begin{figure}[tbp]
\vskip-10mm
\centerline{\psfig{file=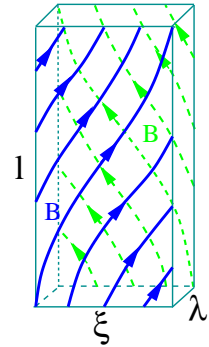,width=3.3in,angle=-90}}
\vskip-10mm
\caption{Anisotropic turbulent eddy in our picture. 
The large-scale magnetic field is in the vertical direction. The field-perpendicular dimensions of the eddy  
are~$\lambda$ and~$\xi\propto \lambda^{3/4}$, and the eddy size in the field-parallel 
direction is~$l\propto \lambda^{1/2}$. 
As the energy cascade proceeds toward the smallest,
dissipative scales,~$\lambda \to 0$, the eddy assumes the
shape of a current sheet.}
\label{beddy}
\end{figure}

To determine the shape of the eddy, we require that the energy flux be constant 
for all scales,~$\delta v_{\lambda}^2/\tau_N={\rm const}$.  
This leads to the scaling of velocity fluctuations $\delta v_{\lambda}\propto (\lambda /\theta_{\lambda})^{1/3}$. 
The displacement of magnetic-field lines is given 
by~$\xi \sim \delta v_{\lambda} l/V_A$, and the correlation length of fluctuations in the field-displacement 
direction cannot be smaller than~$\xi$. Remarkably, the obtained shape of the eddy 
satisfies~$\lambda/\xi \sim \theta_{\lambda}$, so it is indeed 
consistent with the assumed alignment of fluctuations within 
the angle~$\theta_{\lambda}$.  Note that in contrast with the Goldreich-Sridhar 
picture, in our model the eddy is three-dimensionally anisotropic,~$l\gg \xi \gg \lambda$, see Fig.~\ref{beddy}.     

It is natural to assume that turbulent fluctuations are scale invariant, 
which means that~$\theta_{\lambda}$ is a power-law function of~$\lambda$.  
We may parametrize $\theta_{\lambda}\propto \lambda^{\alpha/(3+\alpha)}$, 
which leads to $\delta v_{\lambda}\propto \lambda^{1/(3+\alpha)}$, $\xi\propto \lambda^{3/(3+\alpha)}$, 
$l\propto \lambda^{2/(3+\alpha)}$. We thus obtain that the sole 
requirement of constant energy flux does not define the eddy shape 
uniquely, but leads to a one-parameter  family of solutions.  
The theory is self-consistent for an arbitrary 
parameter~$\alpha\geq 0$. (Note that Goldreich-Sridhar 
model is a particular solution corresponding 
to~$\alpha=0$.) In order to address the crucial  
question about the value of~$\alpha$, we now have to  
use the second conserved quantity -- cross-helicity. In other words, we 
want to find~$\alpha$ that minimizes the total angular 
mismatch between the velocity and magnetic-field polarizations in the eddy.   

\begin{figure}[tbp]
\centerline{\psfig{file=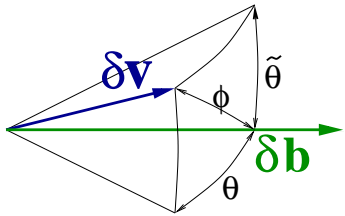,width=1.8in,angle=0}}
\caption{Sketch of three-dimensional angular alignment of shear-Alfv\'en 
velocity and magnetic-field fluctuations. 
The alignment angles consistent with an energy cascade are given 
by~$\theta\propto \lambda^{\alpha/(3+\alpha)}$ 
and~${\tilde\theta}\propto \lambda^{1/(3+\alpha)}$. The maximal alignment is achieved for~$\alpha=1$ (see the text).}  
\label{angles}
\end{figure}

The mismatch angle in the field-perpendicular (horizontal) plane is  
$\theta_{\lambda}\propto \lambda^{\alpha/(3+\alpha)}$. However, the polarization 
vectors are also mismatched in the vertical direction.  
To obtain the vertical alignment angle, 
${\tilde \theta}_{\lambda}$, we note that in the regime of strong turbulence, 
eddies propagating along a large-scale magnetic field interact efficiently  
during only one crossing time. Therefore, only the {\em local} direction of 
the magnetic field matters, 
and when we speak about eddy elongation in the field-parallel direction,~$l$, we should mean 
the eddy dimension along the local magnetic field (this was established by~\citet{cho}). 
It is however important to note that 
the direction of the local magnetic field at the scale~$\lambda$ cannot be defined precisely. 
Since the corresponding eddy contains magnetic field lines wandering within the 
angle ${\tilde \theta}_{\lambda}\sim \xi/l\propto \lambda^{1/(3+\alpha)}$, 
the direction of the local magnetic field can only be defined with the same accuracy.  This means that the 
directions of shear-Alfv\'en velocity-field and magnetic-field fluctuations are aligned 
in the vertical 
direction within the angle~${\tilde \theta}_{\lambda}$, as is sketched in Fig.~\ref{angles}. 
Since both alignment angles, $\theta_{\lambda}$ and $\tilde{\theta}_{\lambda}$, 
are small, the total angular mismatch 
between $\delta {\bf v}_{\lambda}$ and $\pm \delta {\bf b}_{\lambda}$ 
can be calculated as~$\phi_{\lambda}=\sqrt{\theta_{\lambda}^2+{\tilde \theta}_{\lambda}^2}$.  

Following our strategy, we now  
require that the alignment angle $\phi_{\lambda}$ be minimal. We however  
observe that the obtained shape of the eddy precludes us from achieving the perfect 
alignment, $\phi_{\lambda}=0$. Indeed, if for a given small 
scale $\lambda$, we try to maximally align the polarizations in the field-perpendicular (horizontal) 
direction, i.e., to minimize $\theta_{\lambda}\propto \lambda^{\alpha/(3+\alpha)}$, 
we need to set $\alpha\to \infty$. In this case, the fluctuations will be completely 
misaligned in the vertical direction, ${\tilde \theta}\propto \lambda^{1/(3+\alpha)} \sim 1$. 
Similarly, if we try to maximally align them 
in the vertical direction, $\alpha\to 0$, they become misaligned in the horizontal plane. 
This ``uncertainty'' is minimized when~$\theta_{\lambda}\sim {\tilde\theta}_{\lambda}$, in which 
case the maximal angular alignment is achieved and preserved for all scales. 
This determines the scaling parameter uniquely:~$\alpha=1$. The resulting scaling of velocity fluctuations 
is $\delta v_{\lambda}\propto \lambda^{1/4}$, and the field-perpendicular energy spectrum has the form  
\begin{eqnarray}
E(k_{\perp})\propto k_{\perp}^{-3/2}.
\label{en}
\end{eqnarray}
The obtained structure and spectrum of turbulent fluctuations is the main result of this paper. 
~\\

{\bf 3}. {\em Discussion and conclusion.}--- 
It may be reasonable to believe that an external magnetic field is not 
essential for our derivation. Indeed, a local guiding field for small-scale fluctuations 
is naturally provided by large-scale eddies, e.g.,~\citep{maron,milano}. By this analogy,  
the spectrum of isotropic MHD turbulence  
should have scaling~(\ref{en}) as well. We however note that 
to observe this spectrum in numerical simulations of isotropic turbulence  
one would need to reach extremely high resolution 
(to ensure~$\delta b_{\lambda}/\delta b_{\Lambda_0}\ll 1$), 
which is impossible with present-day computer power. 

We also note that our theory naturally explains the presence of ribbon-like dissipative 
structures (current sheets) in numerical simulations of 
MHD turbulence~\citep{biskamp3,maron}. 
Indeed, the form of the eddy predicted in 
our model converges to such a structure as~$\lambda\to 0$.  

On the observational side, MHD turbulence is invoked to explain 
solar-wind measurements, e.g., \citep{goldstein} and  
interstellar scintillation, 
e.g.,~\citep{lithwick}. Although the inferred spectra of magnetic-field and  
electron-density fluctuations are broadly 
consistent with the $-5/3$ scaling, there do exist indications in favour 
of ``$-3/2$'' in some diffractive scintillation~\citep{shishov}.

In conclusion, we propose that similarly to 
decaying MHD turbulence, driven MHD turbulence tends to align the polarizations 
of magnetic- and velocity-field 
fluctuations. However, the dynamic alignment cannot be perfect: 
perfectly aligned fluctuations do not interact and cannot carry energy flux.  
We therefore require that the alignment be maximal 
under the constraint of constant energy flux. Such requirement defines 
the alignment angle uniquely,~$\phi_{\lambda}\propto \lambda^{1/4}$, 
which means that the strength of nonlinear interaction in driven MHD 
turbulence  is reduced by 
the factor $\propto \lambda^{1/4}$ compared to 
a simple dimensional estimate $(\delta v_{\lambda})^2/\lambda$.  
The resulting fluctuations are three-dimensionally 
anisotropic (Fig.~\ref{beddy}), and their energy spectrum is 
$E(k_{\perp})\propto k_{\perp}^{-3/2}$, 
in good agreement with numerical results.

\acknowledgments
I am grateful to Peter Goldreich and Samuel Vainshtein for important  
discussions. This work was supported by the NSF Center for Magnetic
Self-Organization in Laboratory and Astrophysical Plasmas
at the University of Chicago.

\end {document}